# Electronic Textbook as a Component of Smart Kids Technology of Education of Elementary School Pupils

Svitlana Lytvynova [0000-0002-5450-6635]

Institute of Information Technologies and Learning Tools
of National Academy of Educational Sciences of Ukraine,
9 M. Berlyns'koho st., 04060, Kyiv, Ukraine
s.h.lytvynova@gmail.com

**Abstract**. The article sets out to analyze national and foreign experience of use of electronic textbooks in the system of education; to justify the use of Smart Kids technology as a system of methods, forms, and electronic educational game resources, electronic textbooks for educational process in the system of elementary school. Four forms of implementation of Smart Kids technology (Smart Case, Smart Teacher, Smart Class, and Smart Kids) were described considering the facilities of every school as well as the level of information and communication technology qualification of the elementary school teacher. The aim of introduction of the technology for each form of teaching, the necessary equipment, and means for its implementation in elementary school environment were determined. Based on the procedural approach to work of an elementary school teacher, six stages of introduction of the technology were justified. Specific aspects of introduction of blended teaching using the principles of Smart Kids technology were defined. The experience of introduction of electronic textbooks to the system of elementary education of Ukraine was described, the choice of electronic textbooks by elementary school teachers was justified, the comments and suggestions of teachers regarding the arrangement of electronic content in E-textbooks were summarized, the main approaches of teachers to the choice of an electronic textbook and development of their information and communication competence were specified. It was identified that the forms, methods, and techniques of use of electronic textbooks in teaching elementary school pupils require further justification.

## 1    Introduction

The topic of the research is of great relevance due to transition of education to the child-centered, competence building, and activity-based approaches, which is indicated in Conceptual Principles of Secondary School Reform "The New Ukrainian School". Current demands of society to educational process require its practical orientation, in particular use of modern technologies and provision of conditions for self-development and self-expression of pupils, considering their individual characteristics.

The legislative basis for introduction of modern technologies to educational process in the secondary educational establishments is represented by the Order of the Cabinet of Ministers of Ukraine "On approval of the Concept of implementation of "The New Ukrainian School" governmental policy in the sphere of reforming of general secondary education for the period till 2029", the Order of the Cabinet of Ministers of Ukraine "On adoption of the plan of action in introduction of the Concept of implementation of "The New Ukrainian School" governmental policy in the sphere of reforming of general secondary education for the period of 2017 – 2019", the Order of Ministry of Education and Science, Youth and Sports of Ukraine "On measures in introduction of electronic educational content", "On adoption of the Provisions on an electronic textbook".

One of the innovations of elementary school is use of electronic educational resources (EER), namely electronic textbooks (E-textbooks) and electronic educational game resources (EEGR).

As EER we will define educational, scientific, informational, and reference materials and means, developed in electronic format, available on data carriers of any type or on computer networks, rendered via electronic digital technical means, and necessary for efficient organization of educational process, particularly the aspects connected with its provision with the high-quality teaching materials [10].

EER is a component of educational process, having the teaching purpose and being used for learning activities of pupils.

EEGR is a type of electronic educational resource of training purpose which combines cognitive and developing functions, containing holistic theoretical material and competence building tasks on the academic subject, presented in game form [8, p. 133].

E-textbook is an electronic educational edition that comprises systematized training material conforming to the curriculum, involves digital objects of different format, and provides interactivity [9].

In addition to holistic content part, E-textbook contains interactive tasks, namely those competence building; multimedia fragments for illustration of theoretical material, electronic tests for formative assessment, various 3D models, and objects of augmented reality.

Elementary school teachers state that use of E-textbooks and EEGR in elementary school should be carried out due to the technology that considers the age characteristics of pupils, as well as the level of qualification of pupils and teachers in modern educational resources. Scholars note the necessity of development of new models of educational process organization and use of a system of interactive tasks to provide the continuous development of cognitive skills of elementary school pupils and stimulation of their cognition, which requires further investigation.

## 2     Analysis of latest research and publications

E-textbook is the first step in creating a dynamic model of education based on coop-eration and networked learning, simultaneously being adaptive for integration with other Internet tools, including free ones, which provides the conditions for collective formation of new knowledge and implementation of a new concept of education according to social challenges [18].

The works by V.Yu. Bykov [1], S.H. Lytvynova [1], [6], [7], [18], O.M. Melnyk [1], [7], [8], O.O. Rybalko [11] reveal the concept of "EER", specifying the demands to such materials. Gamification aspects intended to enhance the efficiency of teaching pupils at the Mathematics classes in elementary school are investigated by L.O. Zhydilova, K.I. Liashenko, A.L. Stolyarevska [5]. The characteristics and criteria of quality evaluation of E-textbooks are established in the works of foreign scientists [21].

O.H. Yesina and L.M. Lingur theoretically justified that use of electronic textbook in educational process provides development of creative and intuitive thinking; esthetic education by means of graphics and multimedia capacities, development of communicative skills; fostering skills to make an optimal decision [4].

Positive practical results were achieved in defining the characteristics of use of E-textbook for independent work intensification. It was proved that the process of work with E-textbook allows expanding the volume of data and messages submitted, improving the efficiency of extratextual component for self-organization and self-control of academic achievements [13].

Scholars L.L. Bosova and N.E. Zubchenok distinguish positive directions of introduction of E-textbook to educational process, namely additional opportunities for assistance and support of learning activity of every pupil as well as organization and support of group learning activities of pupils [3].

Furthermore, the results suggest two directions in which teachers can use E-textbooks to raise pupils' interest in studying: creation of modern conditions for pupils' work in the classroom and improvement of system of homework preparation [22].

Developing a holistic conceptual vision of an educational edition of a new type, discussing the models of introduction of electronic textbook to common practice, and a range of other crucial aspects are the tasks whose completion largely depend on the work of scientists, practice teachers etc. [14].

Despite a comparative analysis of pupils' choice between printed and E-textbooks, performed by the scholars, no sufficient dependencies in the results of pupils' work or demographic characteristics were established. The simplicity of use of the book was the determining factor of pupils' choice, which should be taken as a basis for development of a modern E-textbook [15; 19].

Some aspects of structure, design, assessment of E-textbook functionality, integration of information and communication technology for the purpose of support of learning were analyzed by scientists from different countries. Nevertheless, some issues of organization of educational process involving E-textbooks were only partially investigated by scholars and thus require additional research [16].

Unresolved aspects of the problem. Despite the experience and results of foreign scientists and certain groundworks on this subject by national scholars, as well as considering novelty of educational processes based on use of textbooks of new type, the questions of developing the models of use of E-textbooks, defining the character-istics of organization of educational process, and efficiency of involving E-textbooks in teaching pupils of general secondary educational establishments of Ukraine were not completely resolved by researchers, which requires further investigation.

The purpose of this article is to justify the use of Smart Kids technology and electronic textbooks in elementary school educational process.

## 3      Methods of Research

The research was conducted as part of investigation "Smart Kids technology of teaching elementary school pupils". The methods used in the course of research include analysis of the theoretical sources, study of the best pedagogical practices of foreign and national specialists in using electronic educational resources and applying them in teaching pupils; synthesis, generalization, and conceptualization to devise the main provisions of the research; modelling the educational process and formulating requirements to electronic textbooks and training of future elementary school teachers; summarization and assessment of results of choice of electronic textbooks by elementary school teachers.

## 4      Research Results

### 4.1     Smart Kids technology of education of elementary school pupils

As K. Hicks states, a methodological subsystem of secondary educational establishments is to be developing constantly. It results from emergence of new technologies in education, namely: introduction of innovations, new approaches to cooperation, cloud technologies, mobile technologies, creation of educational games and gamification of educational process; ubiquitous access to open educational content, monitoring and educational analytics; projecting educational environment [17].

As pedagogical technology we will define a system of methods, forms, and means of conducting any process pertaining to education [2].

*As Smart Kids technology we will define a system of methods, forms, electronic educational game resources and electronic textbooks to provide education of elementary school pupil (Fig. 1).*

*E-textbook and EEGR are the components of Smart Kids technology.*

In general, secondary educational establishments, Smart Kids technology is implemented in the following four forms: Smart Case, Smart Teacher, Smart Class, and Smart Kids. Let us consider them (Fig. 2).

*Smart Case Form.* The aim: use of E-textbooks/EEGR for stimulation of learning activity of pupils in a class. The form of work: teamwork. The necessary equipment: teacher's briefcase with E-textbooks/EEGR, projector, interactive whiteboard, teacher's computer.

*Smart Teacher Form.* The aim: use of E-textbooks/EEGR for ubiquitous access of pupils to teaching materials with the help of their own computers (laptops, tablets). The form of work: frontal and individual. The necessary equipment: teacher's brief-case with E-textbooks/EEGR, projector, interactive whiteboard, teacher's computer, teacher's virtual office, pupils' home computers. Teacher uses the virtual office as an electronic register of quantity and quality of the tasks performed by students. For the purpose of developing an individual pupil's trajectory, a teacher can coordinate task performance by each pupil.

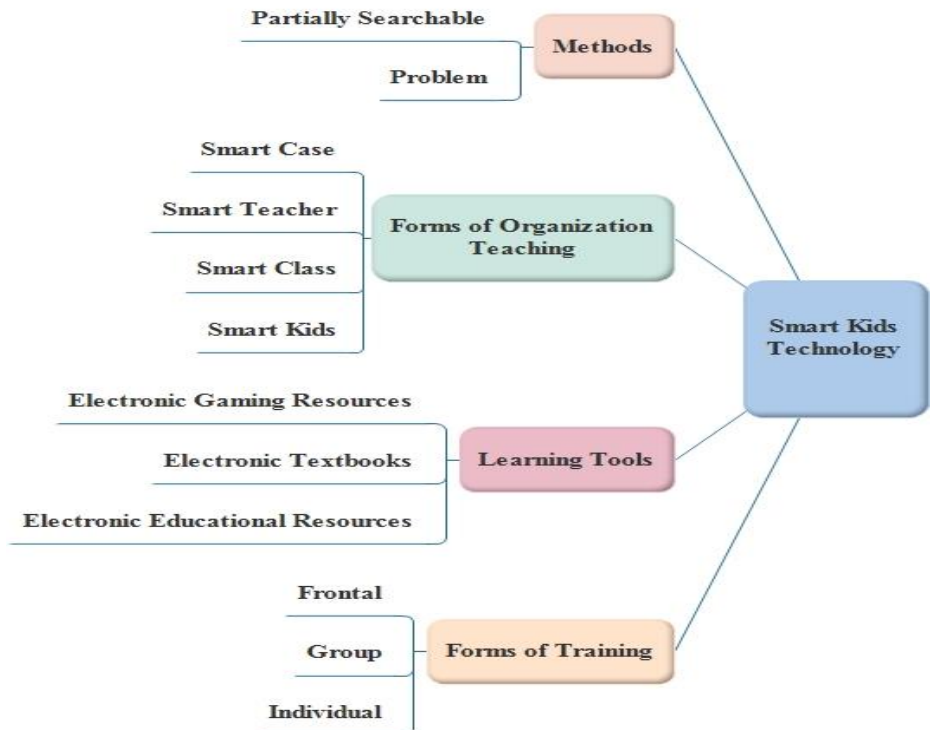

Fig. 1. The Conceptual Structure of Smart Kids Technology

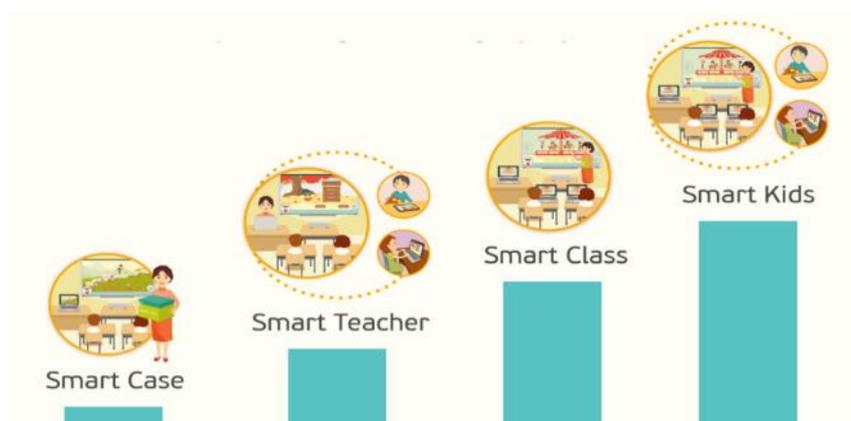

Fig. 2. Forms of implementation of Smart Kids technology in elementary school

*Smart Class Form.* The aim: use of E-textbooks/EEGR for developing an individual pupil's trajectory. The form of work: individual. The necessary equipment: teacher's briefcase with E-textbooks/EEGR, projector, interactive whiteboard, teacher's computer, tablets for each pupil.

*Smart Kids Form.* The aim: use of E-textbooks/EEGR for stimulation of learning activity of pupils in the classroom, provision of the ubiquitous access of pupils to learning materials, and development of an individual pupil's improvement trajectory. The form of work: teamwork, individual, group. The necessary equipment: teacher's briefcase with E-textbooks/EEGR, projector, interactive whiteboard, teacher's computer, tablets for each pupil, teacher's virtual office, pupils' home computers.

Use of E-textbooks should be conducted according to the aim and tasks of a lesson, namely at the beginning, in the middle, at the end of a lesson, or for independent study at home [7, p. 20]. At the beginning of a lesson, a teacher can use E-textbooks/EEGR for refreshing the basic knowledge, checking additional tasks, conducting dictations (mathematical, grammar). For knowledge retention, it's advisable to use E-textbooks in the middle of the class, which will provide the pupils with an opportunity to master the skills in writing or solving problems in game form. At the end of the lesson it is recommended to select the tasks that would allow summarizing the material learned at the lesson.

Methodological characteristics of Smart Kids technology include organization of relay competitions, quests, contests, and crosswords.

To provide a health-preserving approach to studying and comply with the health standards, use of E-textbooks/EEGR at a lesson should not exceed 10-12 minutes.

The main advantages of use of Smart Kids technology include target development of memory, attention, thought, perception of teaching data from the screen, as well as development of culture of use of EEGR and electronic textbooks and competences in educational communication [1], [11], [20].

To introduce Smart Kids technology to elementary school, it's necessary to own E-textbooks/EEGR, while for an elementary school teacher to understand the main stages of introduction of such a technology. Let us consider the stages of introduction of Smart Kids technology to the system of elementary education (Table 1).

**Table 1. Stages of introduction of Smart Kids technology**

| Stage | The content of stage | Procedures |
|---|---|---|
| Stage 1 | Testing one of E-textbooks (or EEGR) during educational process in conditions of elementary school | Mastering skills of integration of EEGR with teaching materials and practical tasks. Use of E-textbooks/EEGR for frontal work with a class. Use of E-textbooks/EEGR in fragmentary mode. |
| Stage 2 | Systematic use of one of E-textbooks (or EEGR) during educational process for frontal work with a class | Building skills of organization of frontal work with pupils (fragmentary mode) |
| Stage 3 | Use of an E-textbook (or | Building skills of organization of group |

|         | EEGR) for learning basic disciplines for group work with a class | work with pupils (fragmentary mode) |
|---------|---|---|
| Stage 4 | Systematic use of an E-textbook (or EEGR) during educational process both for frontal and group work | Organization of efficient frontal and group work with pupils |
| Stage 5 | Systematic use of E-textbooks/EEGR | Establishing pupils' work at home. Introduction of elements of blended teaching |
| Stage 6 | Conscious use of Smart Kids technology for teaching elementary school pupils | Organization of efficient frontal and group work with pupils, blended teaching based on E-textbooks/EEGR |

### 4.2 The aspects of blended teaching of elementary school pupils within Smart Kids technology

Organization of teaching pupils with the help of E-textbooks/EEGR has the features of blended teaching [13] and can be conducted by two methods.

Method 1. While learning in the classroom, one part of a lesson consists in independent work with computers and educational resources, or an electronic textbook, while the other one – in frontal or group work in copybooks.

Method 2. At home, pupils watch video fragments, get familiarized with new topics, perform independent tasks on computer, while in the classroom summarize, master their skills, and acquire competences.

The organization of performing tasks by pupils has some specific characteristics, in particular:

− independent performing of tasks (in the classroom) using a computer, including tablets, under teacher's supervision, complying with health standards (up to 12 minutes) and corresponding to the topic of a lesson;

− independent performing of tasks (at home) using a computer, a tablet, or a smartphone under parents' supervision;

− watching video materials in the classroom; a teacher selects the necessary fragment and presents it to students. Afterwards, the tasks of the following types are performed: to describe the event; to express opinion on the fragment watched; to provide characteristics of a character; to write a continuation etc.;

− watching video materials at home aimed at revision of the material learned, improving knowledge on the topic learned; creating a picture narration with the purpose of development of imaginative thinking or attention.

## 4.3  The electronic textbooks in elementary school

Organization and introduction of Smart Kids technology to the system of elementary education at schools require both means of information and communication technologies (tablets for individual and group work, multimedia means for frontal work) and electronic educational resources, namely electronic textbooks that can be used by pupils during educational process in the classroom and at home.

The Law of Ukraine "On education", adopted in September 2017, specifies that the state guarantees free supply of textbooks (including electronic ones) for applicants for complete general secondary education (p. 5, article 75). This presupposes creation and functioning of a special informational resource on the Internet which will host free complete electronic versions of printed textbooks or electronic textbooks to provide pupils with complete general secondary education.

In 2018, due to "Provisions on electronic textbook", first electronic textbooks for elementary school (1st grade) were developed in Ukraine, including: "Ya doslidzhuiu svit" ("I investigate the world"), "Mystetstvo" ("Art"), "Ukrainska mova. Bukvar" ("Ukrainian language. Primer"), "Matematyka" ("Mathematics") (The Order of the Ministry of Education and Science of Ukraine No. 1078 of 04/10/18 "On assigning the status "Recommended by the Ministry of Education and Science of Ukraine" to electronic textbooks for elementary grades of The New Ukrainian School").

Three publishers – "Ranok", "Heneza", and "Rozumnyky" – took part in development of the textbooks. In selecting the technology of electronic textbook arrangement, "Ranok" and "Rozumnyky" publishers chose the PDF-format of a printed textbook, additionally providing it with a range of interactive exercises, accompanying sounds and developing the navigation similar to that of a traditional one, while "Heneza" publisher followed another approach, creating a textbook by analogy with a computer game.

Organization of the choice of e-textbooks (50 schools of Ukraine taking part in the experiment "Electronic textbook for general secondary education" (EBSE)): e-textbooks were placed in the electronic library; all participants of the experiment had access to them; the choice of e-textbook was recorded in a paper questionnaire and sent for compilation by e-mail. Selection of electronic books by elementary school teachers showed the following results:

1) Electronic textbook on integrated course "Ya doslidzhuiu svit" ("I investigate the world"), 1st grade (Table 2, Fig. 3).

**Table 2. The results of choice of "Ya doslidzhuiu svit" ("I investigate the world") E-textbook**

| No. | Publisher | Authors | Rating |
|---|---|---|---|
| 1 | *"Rozumnyky Publisher" LLC* | Voronstova, T.V., Ponomarenko, V.S., Khomych, O.L., Harbuziuk, I.V., Andruk, N.V, Vasylenko, K.S. | 2/31% |
| 2 | *"Ranok Publisher" LLC* | Bibik, N.M., Bondarchuk, H.P. | **1/37%** |
| 3 | *"Ranok Publisher" LLC* | Bolshakova, I.O., Prystinska, M.S. | 3/26% |

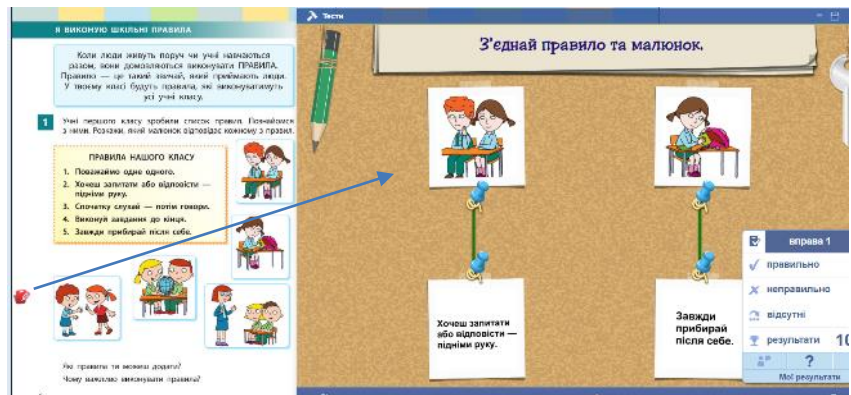

Fig. 3. Interactive exercises of **"I investigate the world" E-textbook**

2) Electronic textbook on integrated course "Mystetstvo" ("Art"), 1st grade (Table 3, Fig. 4).

**Table 3. The results of choice of "Mystetstvo" ("Art") E-textbook**

| No | Publisher | Authors | Rating |
|---|---|---|---|
| 1 | *"Rozumnyky Publisher" LLC* | Kalinichenko, O.V., Arystova, L.S. | 2/29% |
| 2 | *"Heneza Publisher" LLC, "Bristar" PE* | Masol, L.M., Haidamaka, O.V., Kolotylo, O.M. | **1/53%** |
| 3 | *"Ranok Publisher" LLC* | Rublia, T.Ye., Shchehlova, T.L., Med, I.L. | 3/18% |

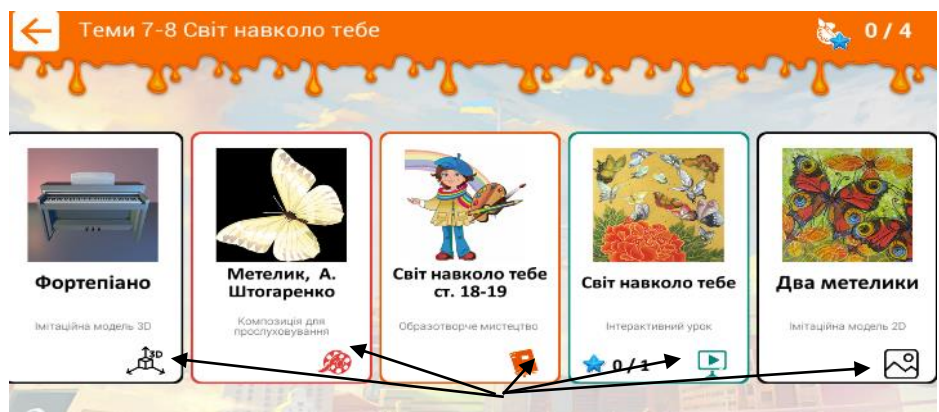

Fig. 4. Interactive exercises of **"Art" E-textbook**

3) Electronic textbook "Matematyka" ("Mathematics"), 1st grade (Table 4, Fig. 5).

**Table 4. The results of choice of "Matematyka" ("Mathematics") E-textbook**

| No. | Publisher | Authors | Rating |
|-----|-----------|---------|--------|
| 1 | *"Rozumnyky Publisher" LLC* | Bevz, V.H., Vasylieva, D.M. | 2/48% |
| 2 | *"Ranok Publisher" LLC* | His, O.M, Filiak, I.V. | **1/52%** |

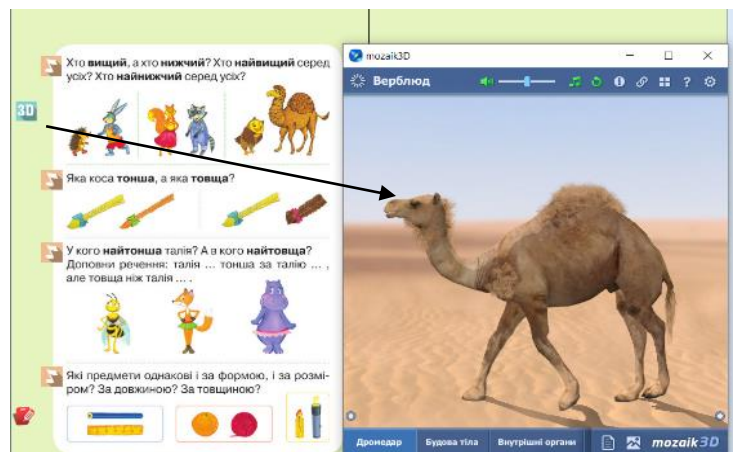

Fig. 5. 3D interactive models in **"Mathematics" E-textbook**

4) Electronic textbook "Ukrainska mova. Bukvar" ("Ukrainian language. Primer") for 1st grade was submitted by one publisher, namely "Rozumnyky Publisher" LLC, the authors: Vashulenko, M.S., Vashulenko, O.V. (Fig. 6).

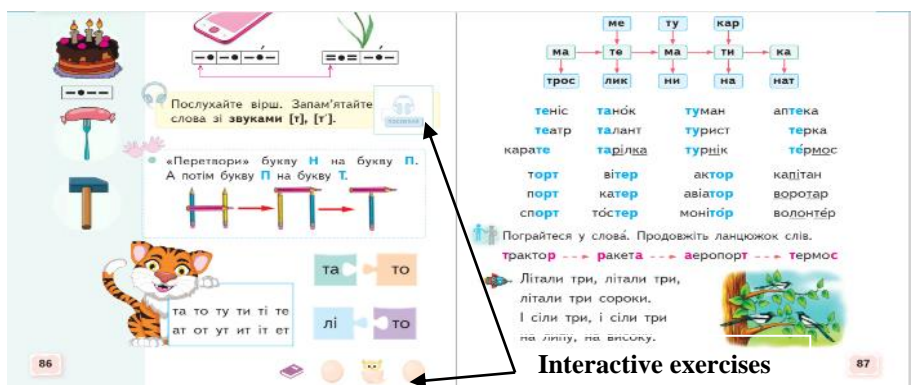

Fig. 6. Text voicing and motivation exercises in **"Ukrainian language: Primer" E-textbook**

The difference in shares of the selected electronic textbooks among the publishers equaled approximately 6%, yet, it was characteristic of only 2 schools.

After selection of textbooks, a blitz survey was conducted, aimed at defining the approaches to the choice and attitude of elementary school teachers to a textbook of new type. Let us consider the results.

*Question 1 related to work experience of the teachers.* The results on work experience of the teachers willing to use first E-textbooks proved to be unexpected. These were not teacher specialists but practice teachers with more than 16 years of work experience (fig. 7).

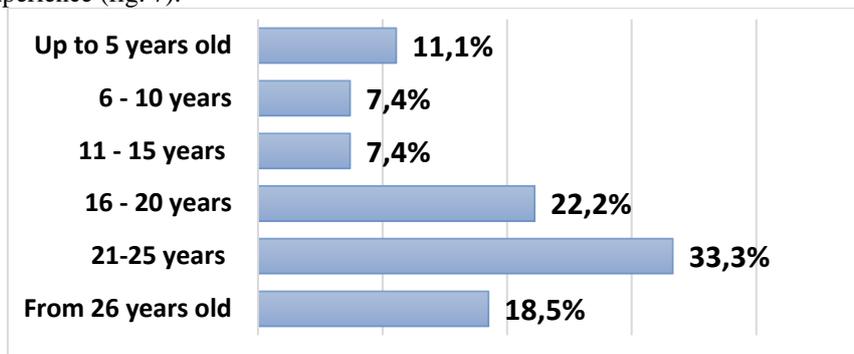

Fig. 7. Work experience of teachers using the first E-textbooks

*Question 2* was connected with approaches to the choice of E-textbooks (fig. 8).

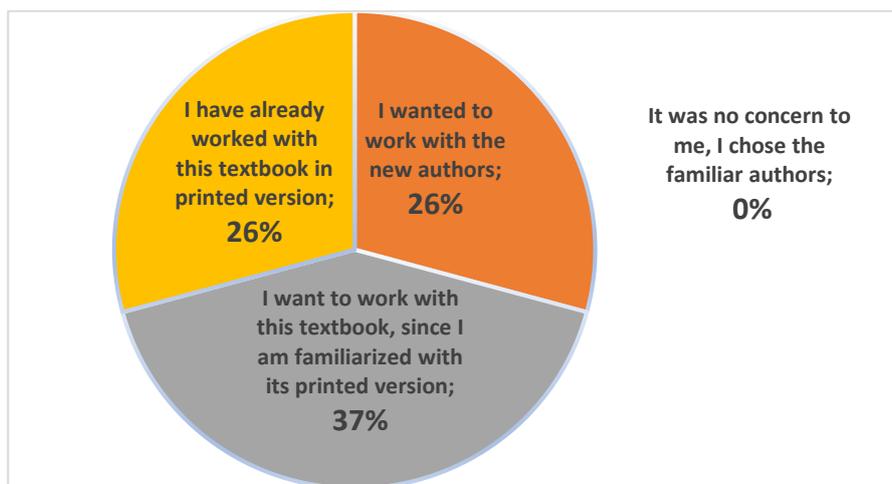

Fig. 8. Approaches to the choice of E-textbooks

As we can see, the majority of teachers – 63% – chose the specific electronic textbook, since they were already familiar with the content of a similar printed textbook, the authors of the textbook, and they had experience of using the printed textbook in educational process.

*Question 3* concerned the positive characteristics of an innovative E-textbook (the process of familiarization) (fig. 9).

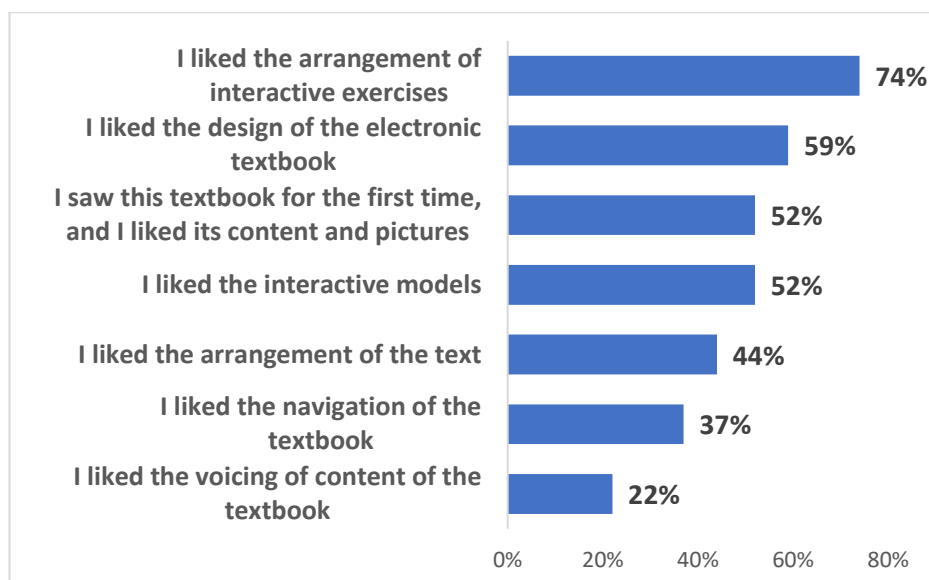

Fig. 9. Positive characteristics of E-textbooks

It would be appropriate to notice that the teachers were satisfied with technologies of arrangement of interactive exercises, content and pictures, the design of E-textbooks, and the interactive models presented.

*Question 4* was dedicated to suggestions of elementary school teachers concerning the content and technologies of representation of E-textbooks. The teachers are of the opinion that it's necessary:
− to increase the number of interactive exercises;
− to enhance the task variability;
− to provide alternation of types of interactive exercises (for diversity at the lessons);
− to add situational exercises, role-play situations, modelling;
− to introduce appendices with the etymology of new words.

In analyzing the answers, it was found that the E-textbook generally satisfies the needs of educational process in elementary school, yet the practice teachers consider it necessary to saturate the E-textbook with interactive exercises (to increase their number) and provide more variability.

Certain comments on the projects of E-textbooks were submitted by The Committee on digital technologies in education with Ministry of Education and Science of Ukraine, in particular:
− cautions on development of electronic textbooks based on paid software;
− certain hyperlinks in E-textbooks were redirecting to resources on YouTube;
− voicing of the text in certain electronic textbooks was partially provided.

All these comments should be considered by publishers in improvement of first electronic textbooks.

Generally, the electronic textbooks presented were highly appreciated by elementary school teachers and, thus, can be used in educational process within elementary education reforming and introduction of pedagogical technologies, namely Smart Kids.

### 4.4 Examples of implementation of E-textbook in teaching 1st grade students

According to the 1st grade curriculum within the New Ukrainian School reform (Ukraine), interactive exercises should correspond to the content of educational material (the designation of sounds by letters; familiarity with letters denoting consonants; teaching the basic technique of reading a direct syllable with a letter designating a vowel sound, etc.) as well as promote students' learning achievements (recognition and distinguishing between letters to indicate vowels and consonants; reading syllables, etc.).

Fragment of the lesson script for 6 years old students (1st grade) using interactive exercises from an electronic textbook (within not more than 12 minutes per a lesson) is represented below.

1. *Start working with an electronic textbook* (Fig. 10).

Students: pronounce the sound and the letter "D", listen to the text, read straight syllables aloud, and work out the skills of navigation and movement of objects in the electronic textbook.

Learning achievements: students get acquainted with the letter "D", study (memorize) its designations, give examples of words with the letter "D" orally, compare sounds.

2. *Work in groups. Didactic game "Recognize the letter D" (Fig. 11).*

Students: perform interactive exercises in the electronic textbook.

Learning Achievements: Students recognize the letter "D" among other letters and count its number.

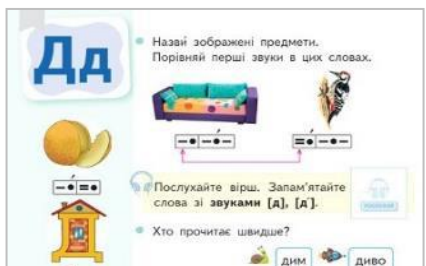 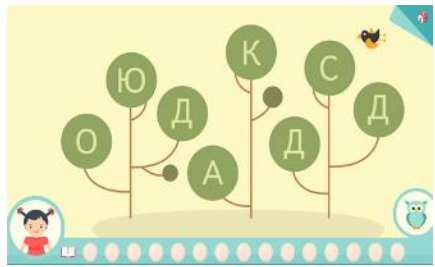

Fig. 10. The fragment of refreshing of students' knowledge        Fig. 11. Fragment of the exercise "Recognize the letter "D""

*3. Individual work (students work on tablets). "Puzzles" didactic game (Fig. 12).*
Students: perform interactive mapping exercises posted in an electronic textbook, pronounce sounds and letters.

Learning Achievements: Students distinguish the letter "D" from the list provided and pronounce the sound and the letter.

*4. Frontal work. Reading syllables in direct and reverse manners (Fig. 13).*

Students: learn syllables, pronounce them, move them, complete interactive exercises with an e-textbook.

Learning Achievements: students acquired the skills of the basic techniques of reading direct syllables with the letter "D", increased the level of IC-competence.

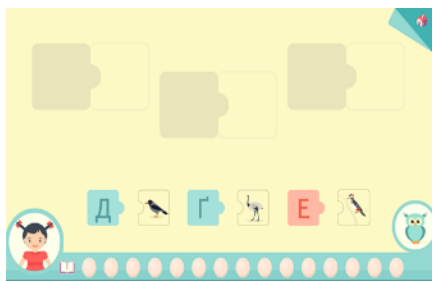 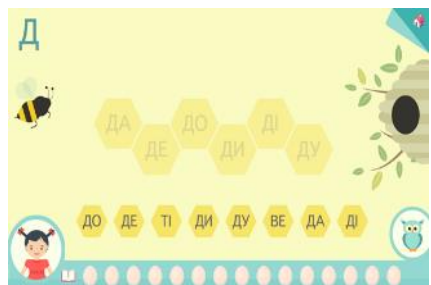

| Fig. 12. The fragment of the interactive exercise "Puzzles" | Fig. 13. The fragment of "Syllables" interactive task |

*5. Let's do some eyes exercises. Finish working with the electronic textbook.*

## 5 Discussing and suggestions on training elementary school teachers in use of Smart Kids technology

Currently, there emerges a need for training elementary school teachers in use of electronic textbooks or EEGR during a lesson, in organization of pupils' work by means of information and communication technologies after-hours, and in supervision of pupils' independent work.

Hence, introduction of Smart Kids technology to elementary school require improvement of the content and including to the development and training program of elementary school teachers the following topics: the concepts of EER, EEGR, and electronic textbook, the structure and development of EER, use of Smart Kids technology for teaching pupils, work with network programs for organization of questioning of pupils.

*The main directions of building competence* of a future elementary school teacher using Smart Kids technology as a component of professional training include: establishment of learning environment of elementary school; use of multimedia complex for work with a class; teacher's briefcase of electronic educational game resources; forms of implementation of Smart Kids technology; methods of teaching students using smart Kids technology; organization and methodological aspects of lesson planning using E-textbooks (namely, EEGR); teacher's virtual office; monitoring pupils' educational attainment; assessment of a lesson using E-textbook (namely, EEGR); assessment of E-textbooks (namely, EEGR).

These directions of training are to be included in curriculums of higher educational establishments specializing in training of future elementary school teachers.

## 6       Conclusions and recommendations for further research

Use of electronic textbooks (or EEGR) diversify the process of teaching, provides transition from passive to active methods of teaching, stimulates learning and cognition activity, allows to develop individual improvement trajectory for each elementary school pupil [8].

Textbooks of new type – electronic textbooks – become an alternative to a traditional printed textbook. Various interactive exercises, voicing, formative assessment provide an opportunity of exciting learning for each child.

New Smart Kids technology provides target development of memory, attention, thought, perception of teaching data from the computer screen, which is crucial for development of personality of a pupil in the 21st century. Smart Kids technology is a system of active target development of a modern elementary school pupil. As a coordinator of this process, a teacher should master the latest teaching technologies based on ICT. Improvement of the system of training of future elementary school teachers require not only basic knowledge of computer equipment, but that of new approaches to establishment of learning environment using new teaching technologies – Smart Kids.

Modern elementary school teachers should master all modern pedagogical and information and communication technologies for the purpose of providing child-centered, competence building, and activity-based studying, as well as implementation of "The New Ukrainian School" Concept.

Forms, methods, and techniques of use of electronic textbooks in teaching elementary school pupils require further justification.